\documentclass[twocolumn]{article}
\usepackage{graphicx}
\usepackage{balance}  

\usepackage[linesnumbered,ruled]{algorithm2e}
\usepackage{subfigure,amsmath,amssymb}
\usepackage{balance} 

\newtheorem{defn}{Definition}[section]
\newtheorem{rem}{Remark}
\newtheorem{exmp}{Example}[section]

\begin{document}

\title{Parallel Clustering of Graphs for Anonymization and 
       Recommender Systems}


\author{
Fr\'ed\'eric Prost \\ 
       {ENS de Lyon, INRIA, UCB Lyon 1, Laboratoire LIP}\\
       {46 Allée d'Italie 69364 LYON Cedex 07, FRANCE}\\
       \texttt{frederic.prost@ens-lyon.fr}
\and Jisang Yoon\\
       {ENS de Lyon, INRIA, UCB Lyon 1, Laboratoire LIP}\\
       {46 Allée d'Italie 69364 LYON Cedex 07, FRANCE}\\
       \texttt{Jisang.Yoon@ens-lyon.fr}
}

\maketitle

\begin{abstract}
  Graph clustering is widely used in many data analysis
  applications. In this paper we propose several parallel graph
  clustering algorithms based on Monte Carlo simulations and expectation
  maximization in the context of stochastic block models. We apply
  those algorithms to the specific problems of recommender systems and
  social network anonymization. We compare the experimental results to
  previous propositions.
\end{abstract}

\section{Introduction}
\label{sec:introduction}
  
An important way to discover structural properties within data is to
classify them \cite{Kle02}, that is to regroup \emph{similar} elements
into classes, called clusters. In this paper we focus on graph data:
for this special case the edges of the graph represent the
correlations between nodes, hence it is the edge topology that is
used to define the clusters of the graph. There are many ways to have
a meaningfull definition of what means to be \emph{similar} (see
\cite{Scha07}).

The stochastic block model, see \cite{SBM1}, is widely used for graph
clustering. The basic idea is that if some nodes behave \emph{similarly} in
network, it can be assumed that those nodes behave the same
\emph{probabilistically} and form a  \emph{cluster} (or a
\emph{block}).

\begin{exmp}
\label{cl:ex0.1}
The simplest stochastic block 
model is depicted in \textbf{Figure \ref{cl:fig1}}.
\begin{figure}[!htp]
\centering
\setlength{\unitlength}{0.8cm}
\begin{picture}(10,4)
\thicklines
\put(4,3){\circle{1}}
\put(3.7,3){$p_{1,1}$}
\put(3,1){\circle{1}}
\put(2.7,1){$p_{2,2}$}
\put(5,1){\circle{1}}
\put(4.7,1){$p_{3,3}$}
\put(3,1){\line(1,0){2}}
\put(2.8,2){$p_{1,2}$}
\put(3,1){\line(1,2){1}}
\put(4.7,2){$p_{1,3}$}
\put(5,1){\line(-1,2){1}}
\put(3.7,1.3){$p_{2,3}$}
\end{picture}
\caption{\label{cl:fig1}3-clustering random graph model}
\end{figure}
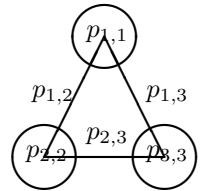

The intuitive meaning of such a model is that $p_{i,j}$ denotes the
probability of having an edge between a vertex in cluster $i$ and a
vertex in cluster $j$.

There is a large class of graphs that fits this model. 
Let us assume that there are $k$-partition of vertices,
$(V_{i})_{i\in [k]}$ (\textbf{Notation}: $[n]:=\{ 0, 1,\cdots ,n\}$
for $n\in \mathbb{N}$). Any graph $G$ such that 
for all $u \in V_{i} , v \in V_{j} (0 \le i \le j < k, u \neq v )$, 
there is an edge between $u , v$ with probability $0 \le p_{i,j} \le
1$ fits the model. 

If we consider the problem the other way around, we have a concrete
graph $G$ for which we look for a SBM of it that is optimal, that is
the optimal clustering  $Z : V (G) \rightarrow [k]$, and parameters 
$p_{i,j}$'s.
\end{exmp}

The goal in \textbf{Example \ref{cl:ex0.1}} is to find the most
\emph{probable} clustering for a given observed network. This kind of
approach is known as the \emph{Maximum Likelihood problem}. Here, the
parameter is the clustering $Z$ (discrete value), and the edge probability
$p_{i,j}'s$ (continuous value).

In the most general case, finding the global optimum is a NP-hard
problem, hence untractable (especially when dealing with very large
data set coming from social networks for instance). However, there
exists a general greedy algorithm (finding the \emph{local} optimum,
iteratively) which is called \emph{Expectation-Maximization (EM)}
algorithm \cite{Dem77} that works well in many problems
\cite{Verb03}. Unfortunalely, there are some difficulties to apply
this algorithm to the problem of graph clustering. In this paper we
suggest a new algorithm to deal with this particular case. To the best
of the authors knowledge, there was no similar algorithm presented in
the literature.

In order to assess the interest of our algorithms, we show that they
work well in practice: both for the quality of the answers and the
time cost. We have considered two case studies: recommender system,
and social network anonymization. For the recommender system case, the
idea is to make clusters of similiar users with relation to
products. For the social network anonymization case, the idea is to
build a SBM of the graph by considering clusters of size $k$. Then an
anonymized graph is created by re-expanding the clusters along the
probabilities of the SBM: we expect to generate a graph with similar
statistical properties of the original one in which it is not possible
to reidentify nodes with an accuracy less than $k$.

We begin in section \ref{sec:expectationmaximization} by recalling
basic definitions and results, and also defining our notations, on the
Expectation-Maximization problems. In section \ref{sec:montecarlo} we
define algorithms for the problem of graph clustering. In section
\ref{sec:applications}, we define more precisely the two domains over
which we have tested our algorithms: recommender systems and social
network anonymization. In section \ref{sec:experimental_results} we
explore the experimental behavior of our algorithms. We discuss
related works in section \ref{sec:related_works}, and finally we
conclude in section \ref{sec:conclusion}.

\section{Expectation-Maximization problems}
\label{sec:expectationmaximization}

Expectation-Maximization (EM) algorithm \cite{DempLaiRub77} denotes a
large class of algorithms to tackle the problem of computing the
maximum likelihood estimates from incomplete data. In this paper we
apply this approach to the problem of graph clustering. We start by
precisely defining the problem setting in its most generic way as well
as our notations.

\begin{defn}[Problem setting]
\label{cl:def1.1}
Let $X$ be the observed data, $Z$ be the unobserved latent
data (in the case of discrete values we write $\mathcal{Z}$ the set of
possible values of $Z$), and $g$ be the probability mass function (pmf) of
$Z$. Thus we derive $g(Z)=\int f_{\theta}(X,Z)$, where $\theta$ is a
parameter of the distribution and $f_{\theta}(X,Z)$ is the probability
density function (pdf) or pmf of $X,Z$.

The goal is to maximize the log-likelihood function, that is to
find 
\begin{equation} \hat{\theta}:=\underset{\theta}{argmax} \text{ }
  f_{\theta}(X)
\end{equation} where $f_{\theta}(X)$ is defined as \begin{equation}
f_{\theta}(X):=\sum_{Z\in \mathcal{Z}}f_{\theta}(X,Z)
\end{equation}
\end{defn}

\indent For computational convenience, we usually take the log to the
likelihood function $f_{\theta}$, and since such value is negative, we
define the \emph{entropy} as follows: 
\begin{equation}
  S(\theta)=-log\circ f_{\theta}
\end{equation}

 The entropy is used since we we may not be able to express 
$f_{\theta}$ as a float value  for very large values  of 
$f_{\theta}$, e.g. $f_{\theta}\approx 10^{10^{6}}$.

\indent Thus, from this point, our goal is to minimize entropy,
instead of maximizing the log-likelihood function. The generic EM
algorithm, for which local convergence is proved in \cite{wu1983}, 
can be written as follows:
\begin{algorithm}[!ht]
\SetKwInOut{Input}{Input}
\SetKwInOut{Output}{Output}
\Input{observed data $X$, unobserved data $Z$, pmf $f_{\theta}(X,Z)$}
\Output{optimal $\hat{\theta}$}
Pick random $\theta$\\
\Repeat{$S(\theta)$ converges}
       { $g(Z) \gets g(Z | X,\theta):=\frac{f_{\theta}(X,Z)}
            {\underset{Z'}{\sum}f_{\theta}(X,Z')}$~~(E-step)\\
          $\theta \gets \underset{\theta}{argmax}\text{ }
            \mathbb{E}_{g}[log\circ f_{\theta}(X,Z)]$~~(M-step)
}
\caption{Generic EM}
\end{algorithm}\\

\begin{exmp}[$k$-means]
\label{cl:ex1.3}

Regarding the way we consider the unobserved latent variable $Z$ there
are two types of EM: soft EM and hard EM. If we consider $Z$ as a
\emph{variable}, we are in the case of a soft EM problem.  If we
consider $Z$ as \emph{parameter}, which we want to figure out, we are
in the case of a hard EM problem. In hard EM, we assume that $Z$ has a
deterministic value.

The generic algorithm of a hard EM is represented as \textbf{Algorithm
  \ref{cl:alg2}}.

\begin{algorithm}[!ht]
\SetKwInOut{Input}{Input}
\SetKwInOut{Output}{Output}
\Input{observed data $X$, unobserved latent data $Z$, pdf or 
pmf $f_{\theta,Z}(X)$}
\Output{optimal $\hat{\theta}$,$\hat{Z}$}
Take random $\theta$\\
\Repeat{$S(Z)$ converges}{
$Z \gets \underset{Z'}{argmin}\text{ }S(\theta ,Z')$(E-step)\\
$\theta \gets \underset{\theta'}{argmin}\text{ }S(\theta',Z)$(M-step)
}
\caption{\label{cl:alg2}Generic Hard EM}
\end{algorithm}

$k$-mean clustering is a particular case of hard EM. For this problem
$X \subseteq \mathbb{R}^{d}$ is an unclassifed data set, and $Z : X
\rightarrow [k]$ is a function which corresponds to a
classification. The parameter $\theta$ is a $k$-tuple $(\mu_{i}\in
\mathbb{R}^{d})_{i\in [k]}$. The probability function $f_{\theta,Z}$
can be defined as follows:
\begin{equation}
  f_{\theta,Z}(X ):=\prod_{x\in X}(norm\_dist(\mu_{Z(x)},I_{d},x))
\end{equation}
where $norm\_dist(\mu, \Sigma, x)$ is defined as the pdf of
$\mathcal{N}(\mu, \Sigma)$. Thus in line 3 of \textbf{Algorithm
  \ref{cl:alg2}}, for each $x \in X$, $Z(x)$ is assigned to
$\underset{i\in [k]}{argmin}|x - \mu_{i} |^{2}$ , and in line 4,
$\mu_{i}$ is computed as the euclidean center of cluster $Z^{-1}(\{ i \})
\subseteq X$. Simply sayed, $k$-mean clustering is a hard EM for a
Gaussian mixture model with fixed covariance.
\end{exmp}

\section{Graph Clustering}
\label{sec:montecarlo}

\subsection{Soft graph clustering (soft SBM)}
\label{cl:sec2.1}

We work on the problem of soft clustering for various kinds of
graphs including simple graphs, weighted graphs \emph{etc}. Each edge
is assumed to have some configuration $r\in R$ ($R$ is the set
of configurations, \emph{e.g.} existence, weight, rating, \emph{etc}).
Let us precisely define the problem setting of soft SBM.

\begin{defn}[Soft SBM]
\label{cl:def2.1}
let $X$ be the observed data (like the existence, the weight, the
rating, \emph{etc.}) of a given set of edges of a graph $G=(V,E)$. Let 
 $h=(h_{u,i})_{u\in V(G), i\in [k]}$, where $\underset{i\in
   [k]}{\sum}h_{u,i}=1,\forall u\in V(G)$, and
 $\theta=(\theta_{i,j,r})_{i,j\in [k],r\in R}$, where $\underset{r\in
   R}{\sum}\theta_{i,j,r}=1,\forall i,j\in[k]$. 
Then the goal function $S(\theta,h)$ is defined as 
\begin{equation}
  S(\theta,h):=-log\circ \prod_{(u,v)\in X}(\sum_{i,j\in
    [k]}h_{u,i}\cdot h_{v,j} \cdot \theta_{i,j,r(u,v)})
\end{equation} 
where $r(u,v)$:=config of $(u,v)$(which is observed). The aim is to
minimize $S(\theta,h)$.
\end{defn}

\begin{rem}
  $(\theta,h)$ corresponds to $\theta$ in \textbf{Definition
    \ref{cl:def1.1}}. In addition, $Z$ is defined as a function from
  $X$ to $[k]\times [k]$, and $f_{\theta,h}(X,Z)$ is defined
  as: \begin{equation} f_{\theta,h}(X,Z)=\prod_{(u,v)\in
      X}(h_{u,i}\cdot h_{v,j}\cdot \theta_{i,j,r(u,v)})
\end{equation} where $Z(u,v)=(i,j)\in [k]^{2}$ for each $(u,v)\in X$.
\end{rem}

\cite{GLGMS16} previously suggested an algorithm (\textbf{MMSBM},
Mixed Membership SBM) based on EM. Here, we propose a
\emph{randomized} algorithm (\textbf{MCMMSBM}, Monte-Carlo MMSBM), for
improving the complexity of \textbf{MMSBM}. The algorithm is based on
a Monte-Carlo simulation and \textbf{MMSBM}. We discuss more precisely
  the relation between \textbf{MCMMSBM} and \textbf{MMSBM} in the
  related works section (see section \ref{sec:related_works}).

\begin{algorithm}[!ht]
\SetKwInOut{Input}{Input}
\SetKwInOut{Output}{Output}
\Input{observed edge data $X$, $V(G)$:set of vertices, $k$: number 
       of clusters, $s$: sample size}
\Output{optimal $\hat{\theta},\hat{h}$}
Take random $\theta$, $h$\\
\Repeat{$S(\theta,h)$ converges}{
\ForPar{$(u,v)\in X,i\in [k]$}{
$x_{u,v}(i),x_{v,u}(i)\gets 0$
}
\ForPar{$i,j\in [k],r\in R$}{
$\eta_{i,j,r}\gets 0$
}
\ForPar{$(u,v)\in X$}{
$isample\gets discrete(h_{u},s)$, $jsample\gets discrete(h_{v},s)$\\
$x\gets 0$\\
$r\gets r(u,v)$\\
\For{$s' \in [s]$}{
$i\gets isample[s'],j\gets jsample[s']$\\
$x,x_{u,v}(i),x_{v,u}(j)+=\theta_{i,j,r}$
}
\For{$s' \in [s]$}{
$i\gets isample[s'],j\gets jsample[s']$
$x_{u,v}(i),x_{v,u}(j)/=x$\\
$\eta_{i,j,r}+=\theta_{i,j,r}/x$
}
}
\ForPar{$u\in V(G), i\in [k]$}{
$h_{u,i}\gets \frac{1}{|\partial u|}\sum_{v\in\partial u}x_{u,v}(i)$
}
\ForPar{$i,j\in [k],r\in R$}{
$\theta_{i,j,r}\gets \frac{\eta_{i,j,r}}{\sum_{r' \in I}\eta_{i,j,r'}}$
}
}
\caption{\label{cl:alg4}MCMMSBM}
\end{algorithm}

$\partial u$ is defined as $\{v\in V(G) | (u,v) \in X \}$. The time
complexity of \textbf{MCMMSBM} is $O(|X|\cdot (k +s\cdot log(k)))$.
The sample size $s$ is chosen by trade-off between cost and
accuracy. Note that the time cost to generate $s$ samples from an
arbitrary discrete distribution with size $k$ is $O(k+s\cdot
log(k))$. It can be implemented by binary search in cumulative
probabilities.\\

\indent Experimental results (see \textbf{Section
  \ref{sec:experimental_results}}) show that \textbf{MCMMSBM} 
and \textbf{MMSBM} produce similar results in terms of quality, 
but that \textbf{MCMMSBM} requires less resources than \textbf{MMSBM}.

\subsection{Hard graph clustering}
\label{cl:sec2.2}

\subsubsection{Hard classification problems and hard EM }

Let us start by defining the problem setting of hard
classification as follows:
\begin{defn}
  $X$: observed data, $A$: unclassified data, $Z$: unobserved latent
  classification from $A$ to $[k]$, $\theta$: parameter distribution,
  $f_{\theta,Z}(X)$: pmf of $X$. The goal is to 
  minimize \begin{equation} S(\theta,Z):=-log\circ f_{\theta,Z}(X)
\end{equation}
\end{defn}

For given $Z$, computing optimal
$\hat{\theta}:=\underset{\theta'}{argmin}\text{ }S(\theta',Z)$ is
usually not very expensive. Indeed, we can differentiate the entropy
function for $\theta$, since $\theta$ has continuous value while $Z$
doesn't. Thus, in hard classification problems, we only consider $Z$
as a parameter, and we can express entropy as 

\begin{equation}
  S(Z):=\underset{\theta'}{min}\text{ }S(\theta',Z)
\end{equation}

The $k$-means algorithm (Gaussian mixture with fixed covariance) is
also an algorithm for hard classification problems (\textbf{Example
  \ref{cl:ex1.3}}). In the most general case (see the line 3 of
\textbf{Algorithm \ref{cl:alg2}}), the problem to solve boils down to
the computation of such an \emph{argmin} in general classification
problems.  The number of possible $Z$ is equal to
$|\mathcal{Z}|=k^{|A|}$. Thus, it is untractable to compare all the
$k^{|A|}$ cases.

The Gaussian mixture model is just a special case in which we can
easily compute $argmin$ in line 3. Indeed, one can just compute argmin
$\underset{Z(a)}{argmin}f_{\theta ,X} (Z)$ for each $a \in A$.

\indent However, computing the clustering $Z$ for a given parameter
$\theta$ is not easy. Indeed, deciding the best $Z(v)$ for each $v \in
V (G)$ \emph{depends} on the values of $Z$ for others $v \in V
(G)$. Thus, we can't decide $Z(v)$, separately as it is the case in a
Gaussian mixture model. Hence, we proposed a new classification
algorithm, \emph{Generalized $k$-means}, which is inspired from the
original $k$-means algorithm.
\begin{algorithm}[!ht]
\SetKwInOut{Input}{Input}
\SetKwInOut{Output}{Output}
\Input{observed data $X$, $A$ : set of unclassified data, $k$ : number
  of clusters, $\alpha \in (0,1]$ : iteration scale}
\Output{$Z$ which minimizes $S(Z)$}
Take random $Z$\\
\Repeat{$S(Z)$ converges}{
$A' \gets sample(A,\alpha |A|)$(uniform sample)\\
\ForPar{$a \in A'$}{
$i\gets \underset{i'\in [k]}{argmin}\text{ }S(succ(Z,a,i'))$\\
Plan to reassign $Z(a)\gets i$
}
Do plan, compute some information with $X,Z$(\emph{e.g.} $\hat{\theta}$)
}
\caption{\label{cl:alg5}Generalized $k$-means}
\end{algorithm}\\

\indent Let $succ(Z, a, i) : A \rightarrow [k]$ be defined as a new
classification satisfying:

\begin{equation}
succ(Z,a,i')(a'):=\begin{cases}
Z(a') &\text{if $a'\neq a$}\\
i' &\text{else}
\end{cases}
\end{equation}
\begin{defn}
If no single movement between two clusters can improve the entropy of
classification, then we call the such a classification a \emph{locally
  optimal classification}. 
\end{defn}

Generalized $k$-means algorithm achieves locally optimal
classification, as same as classical $k$-means algorithm.

To compute $S(Z)$, we have to compute $\hat{\theta} (Z)$, and we can
approximate 
\begin{equation}
  \begin{split}
   S(succ(Z,a,i'))&:=-log(f_{X,succ(Z,a,i')}(\hat{\theta}(succ(Z,a,i'))))\\
   &\approx -log(f_{X,succ(Z,a,i')}(\hat{\theta}(Z)))
  \end{split}
\end{equation} 
if we assume that $|A|\gg 1$, and MLE of $f_{X,Z} (\theta )$ are
\emph{consistent}, this approximation works very well. 

The third line of \textbf{Algorithm \ref{cl:alg5}} is the key part of
the algorithm. If $\alpha$ is set to $1.0$, then this algorithm may not work
for the hard graph clustering problem. Indeed, in this case it is not
possible to assume that the proportion of movement during the
parallel-loop is \emph{negligible}.

\begin{rem}
 If $S(Z)$ starts to decrease in early iterations, it may continue to
 decrease (hence converge), because as $S(Z)$ decreases, most of $Z(a)$ 
 for $a \in A'$ may keep its value, so that $\underset{i'\in
    [k]}{argmin}S(succ(Z,a,i'))$ becomes more accurate.
\end{rem}

\subsubsection{Hard clustering of simple graph}

Let us consider a real network as an observed data from a random graph
model with clustering. Actually, such an approach constitutes a random
graph model for \emph{simple graphs}. It is possible to extend this
approach to other kinds of graphs (\emph{e.g.} directed graphs,
weighted graphs, \emph{etc.}).

Now, let us apply Generalized $k$-means on a hard simple graph
clustering problem.

First, let us compute $S(Z)$. 
\begin{equation}
  \label{cl:eq21}
  \begin{split}
    S(Z)=&-\sum_{0\le i \le j < k}(d_{i,j}log(\hat{p}_{i,j})+
           d'_{i,j}log(1-\hat{p}_{i,j}))\\
        =&\sum_{0\le i \le j < k}((d_{i,j}+d'_{i,j})log(d_{i,j}+d'_{i,j})\\
        &-d_{i,j}log(d_{i,j})-d'_{i,j}log(d'_{i,j}))\\
        &\text{(because MLE of Bernoulli distribution)}\\
        =&\sum_{0\le i \le j < k}f(d_{i,j},d'_{i,j})
   \end{split}
\end{equation} where \begin{equation}
\label{cl:eq22}
\begin{split}
  &V_{i}:=Z^{-1}(\{ i \})\\
  &D_{i,j}:=
  \begin{cases}
    |V_{i}|(|V_{i}|-1)/2 &\text{if $i=j$}\\
    |V_{i}||V_{j}| &\text{else}
  \end{cases}\\
  &d_{i,j}:=\text{($\#$ edges between $V_{i}\& V_{j}$)}\\
  &d'_{i,j}:=D_{i,j}-d_{i,j}\\
  &f(x,y)\\
  &:=\begin{cases}
       (x+y)log(x+y)-x\cdot log(x) -y\cdot log(y) &\text{if $x,y \ge 0$}\\
       0 &\text{else}
     \end{cases}
\end{split}
\end{equation}

Now, let us consider \textbf{Algorithm \ref{cl:alg8}}.
\begin{algorithm}[!ht]
  \SetKwInOut{Input}{Input}
  \SetKwInOut{Output}{Output}
  \Input{$G$: Observed simple graph data, 
         $Z:V(G) \rightarrow \{1,2,\cdots ,k \}$: unobserved latent 
            classification data, 
         $\alpha \in (0,1]$: iteration scale }
  \Output{$Z$ which minimizes $S(Z)$}
  Take random $Z$\\
  \Repeat{$S(Z)$ converges}{
    $V' \gets sample(V(G),\alpha |V(G)|)$(uniform sample)\\
    \ForPar{$v \in V'$}{$i\gets \underset{i'\in [k]}{argmin}
            \text{ }S(succ(Z,v,i'))$(\textbf{Algorithm \ref{cl:alg10}})\\
            Plan to reassign $Z(v)\gets i$}
     Do plan, compute
     $d_{i,j},d'_{i,j},x_{v,j},A_{i},a_{i,j},B_{i},b_{i,j}$
     (\textbf{Algorithm \ref{cl:alg9}}) 
     for $0\le i \le j < k,v\in V(G)$
  }
\caption{\label{cl:alg8}Simple graph clustering via Generalized k-means}
\end{algorithm}
\begin{algorithm}[!ht]
  Compute $D_{i,j}, d_{i,j},d'_{i,j},V_{i}$ as in Eq (\ref{cl:eq22})\\
  $x_{v,j}\gets \text{($\#$ edges from $v$ to $V_{j}$)}$\\
  \ForPar{$i\in [k]$}{
    $A_{i}\gets 0$\\
    $B_{i}\gets 0$\\
    \For{$1 \le j \le k$}{
      $a_{i,j}\gets f(d_{i,j},d'_{i,j}+|V_{j}|)-f(d_{i,j},d'_{i,j})$\\
      $b_{i,j}\gets f(d_{i,j},d'_{i,j})-f(d_{i,j},d'_{i,j}-|V_{j}|)$\\
      $A_{i}\gets A_{i}+a_{i,j}$\\
      $B_{i}\gets B_{i}+b_{i,j}$
    }
  }
\caption{\label{cl:alg9}substep of \textbf{Algorithm \ref{cl:alg8}}}
\end{algorithm}

\begin{algorithm}[!ht]
  $temp\gets B_{y}$\\
  \For{$j$ s.t. $x_{v,j}\neq 0$}{
    $temp\gets temp-b_{y,j}$\\
    $y_{v,j} \gets |V_{j}|-x_{v,j}$\\
    $temp\gets temp+f(d_{y,j},d'_{y,j})-f(d_{y,j}-x_{v,j},d_{y,j}-y_{v,j})$
  }
  $min \gets temp$\\
  $argmin\gets y$\\
  \For{$y\in [k]$}{
    $temp\gets A_{y}$\\
    \For{$j$ s.t. $x_{v,j}\neq 0$}{
      $temp\gets temp-a_{y,j}$\\
      $y_{v,j}\gets |V_{j}|-x_{v,j}$\\
      $temp\gets temp+f(d_{y,j}+x_{v,j},d_{y,j}+y_{v,j})-f(d_{y,j},d'_{y,j})$
    }
    \If{$temp < min$}{
      $min \gets temp$\\
      $argmin \gets y$
    }
  }
\caption{\label{cl:alg10}substep of \textbf{Algorithm \ref{cl:alg8}}}
\end{algorithm}

$A_{i}$ represents the entropy increment corresponding to the merging
of an isolated vertex to the cluster $V_{i}$.  $B_{i}$ represents the
entropy decrement corresponding to the split of an isolated vertex
from the cluster $V_{i}$. The reason for the computations of $A_{i},
B_{i}$ is that the real network is seen as a sparse network. Thus, we
can consider any arbitary vertex $v\in V'$ in \textbf{Algorithm
  \ref{cl:alg8}} as an \emph{almost} isolated vertex. If there are
some edges from $v$ to $V_{j}$, one just has to modify $A_{i}$. Then,
the time complexity of line 5 in \textbf{Algorithm \ref{cl:alg8}} is
improved from $O(k^{2})$ to $O(min(d\cdot k,k^{2}))$, where $d$ is the
average degree($=\frac{2m}{n}$). If $O(m)\approx O(n)$, then $d$ is a
constant, thus the complexity for line 5 is $O(k)$.

Now, let's compute the global complexity of \textbf{Algorithm
  \ref{cl:alg8}}. $\alpha \cdot s$ can be considered as the number of
repeat-loop iterations, because all the vertices in $V(G)$ have to be
correctly assigned to the clusters. Let $m, n$ be the respecrtive
numbers of edges and vertices in $G$. The for-loop in line 4 may take
$O(min(m \cdot k,n\cdot k^{2}))$ by \textbf{Algorithm \ref{cl:alg8}},
line 8 take $O(m+k^{2})$, but by considering $m \gg k^{2}$, we have
$O(m+k^{2})\approx O(m)$. In conclusion, the complexity of
\textbf{Algorithm \ref{cl:alg8}} is $O(\frac{min(m\cdot k,n\cdot
  k^{2})}{P})$, where $P$ is the number of processors.

\section{Applications}
\label{sec:applications}

  \subsection{Recommender System}
  \label{subsec:recommander}

\begin{figure}[!ht]
\centering
\includegraphics[scale=0.08]{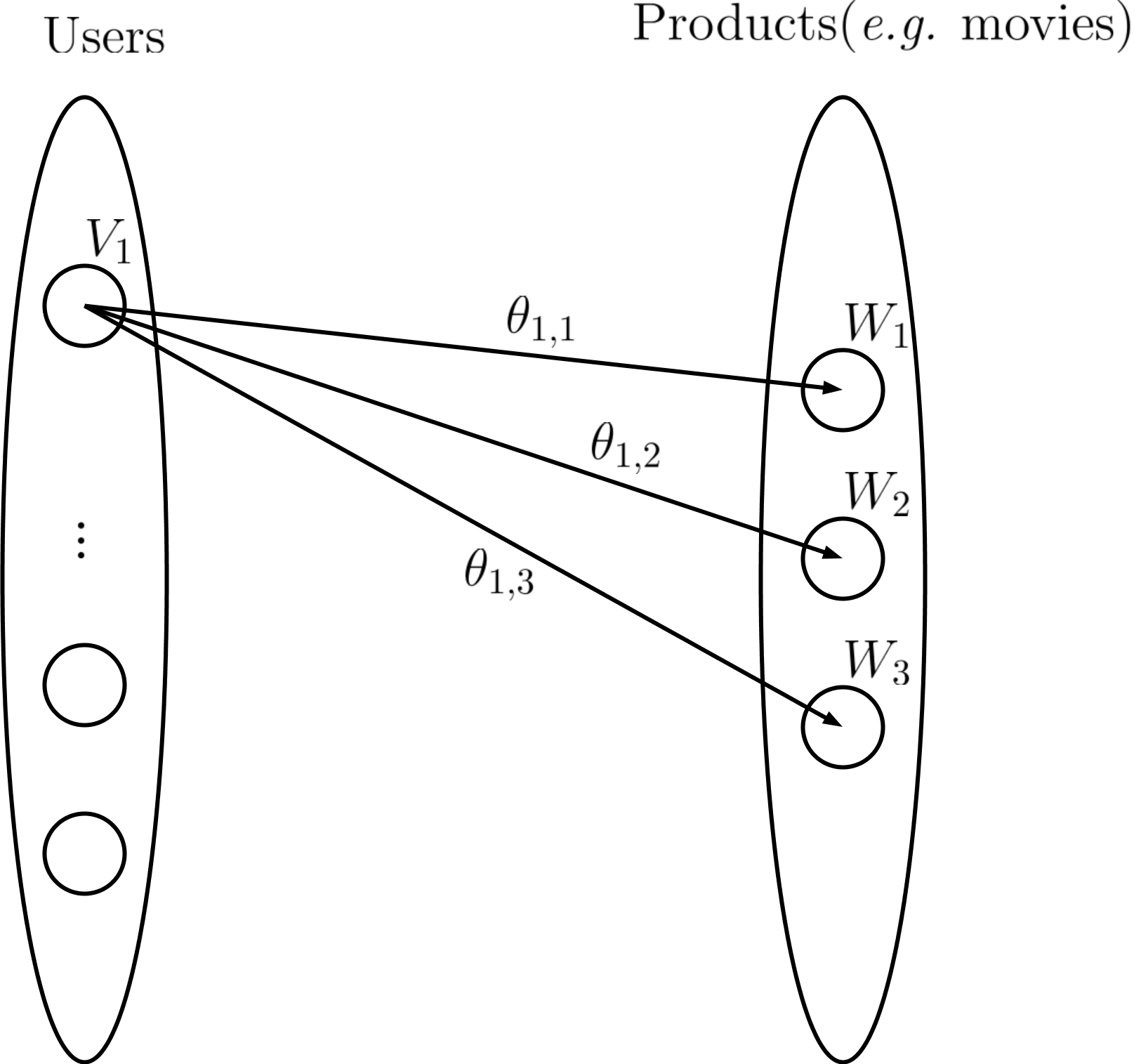}
\caption{\label{cl:fig3}SBM for recommender system}
\end{figure}

  \cite{Boba13} One of the basic approach to recommender system is to
  consider the user-product relationship as a bipartite graph (see
  \textbf{Figure \ref{cl:fig3}}). Users may rate each product they
  purchase or press the \emph{like} button on some products. The
  former case can be considered as a weighted graph, and the latter
  case as a simple graph. The observed data is not the full graph,
  because each user might not experience/purchase all the
  products. Usually only a \emph{partial} observation of the edges of
  the graph is considered. The goal is to anticipate the weight or the
  existence of hidden (unobserved) edges.

\indent In order to anticipate the weight or the existence of the
edges, we look at this problem through the SBM point of view. Let us
consider that there are groups of \emph{similar users or similar
  products}. The existence or the weight of edges between clusters
$V_{i},W_{j}$ follow some random distribution model (\emph{e.g.}
Bernoulli, binomial, \emph{etc.}, see \textbf{Figure
  \ref{cl:fig3}}). Using \textbf{MCMMSBM}, \textbf{MMSBM}, and
Generalized $k$-means, the clustering can be optimized, and used to
anticipate the weight or the existence of the hidden edges.

  \subsection{Social Network Anonymization}
  \label{subsec:anosoc}

  The information in social networks becomes an important data source,
and sometimes it is necessary or beneficial to release such data to
the public. Many real-world social networks contain sensitive
information and serious privacy concerns on graph data have been
raised. The famous result of Narayan and Shmatikov \cite{NarayanS09}
has shown that na\"{\i}ve anonymization does not work: it is in practise
very easy to re-identify elements of a trivially anonymized (ie
replacing identifying informations such as names, social security
numbers etc. with random numbers) social networks. Later works
\cite{BackstromDK11} pushed further the study of attacks on
anonymized social networks.

  The goal of social network anonymization is to produce a graph in
such a way that some statistical functions produce the same result on
the original graph and on the transformed graph, while other functions
(namely reidentification) should not produce the same result. There
are two main ways to work on the anonymization:
\begin{enumerate}
\item Clustering: one tries to group together edges and nodes so that
  when the the cluster regroups $k$ nodes then there is no way
  to distinguish an individual node among them.
   \item k-anonymity: one tries to modify the original graph in such a
     way that there should be at least k-1 other candidate nodes
     with similar features (the features are part of the assumption
     made on the capability of the attacker). 
\end{enumerate}

 In both cases one can assure that re-identification cannot be more
precise that randomly picking among at least $k$ candidates. It looks
natural to apply our algorithms to the clustering approach 
(actually Hay \emph{et al}(\cite{hay10}) suggested a similar approach
for social network anonymization see section
\ref{sec:related_works}).

\section{Experimental Results}
\label{sec:experimental_results}
\indent Programs have been implemented with C++, OpenMP. The hardware
configuration for experiments is given in \textbf{Table
  \ref{par:hardware}}.
\begin{table}[!ht]
\centering
\begin{tabular}{| l | c |}
\hline
Name & Crunch1\\
\hline
CPU & 4 x Intel(R) Xeon(R) \\
 & CPU E5-4620 0 @ 2.20GHz\\
\hline
Total cores & 32\\
\hline
Memory(GB) & 379\\
\hline
System type & PowerEdge R820 (Dell Inc.)\\
\hline
\end{tabular}
\caption{\label{par:hardware}Hardware configuration}
\end{table}
\subsection{Comparison between MMSBM and \\MCMMSBM}
\label{cl:sec6.1}
\begin{figure}[!ht]
\centering
\includegraphics[scale=0.4]{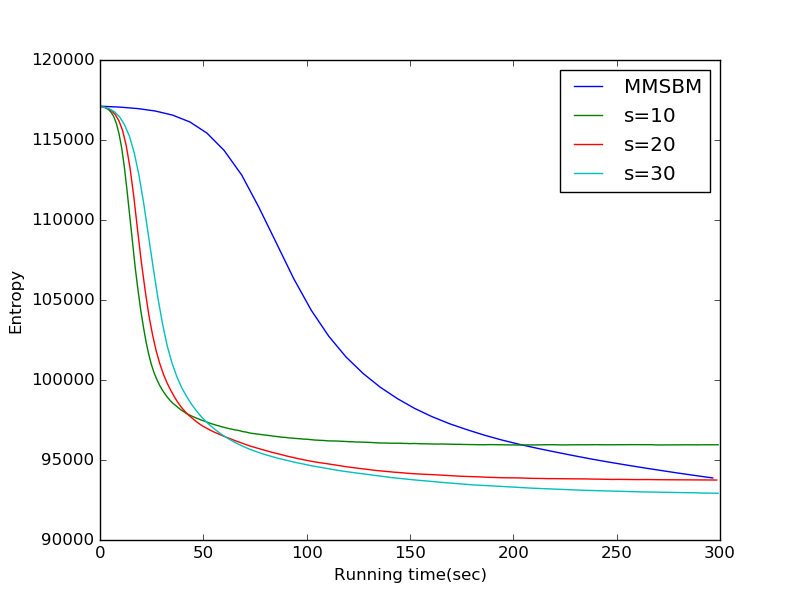}
\caption{\label{cl:fig4}}
\end{figure}

The benchmark for the comparison between \textbf{MMSBM} and
\textbf{MCMMSBM} the recommender system (see \textbf{Section
  \ref{sec:applications}}). The theoretical basis of both
\textbf{MMSBM} and \textbf{MCMMSBM} is similar same. The algorithms
are slightly different. Both are using soft clustering, here, we
assume that rating between two clusters $U_{i}$, and $V_{j}$ follows
\emph{multinomial} distributions. we used the 100k movielens dataset
\cite{movielens1}.  For the evaluation of the predictions, 5-fold
cross validation are done. First, let's compare how these algorithms
optimize entropy. Results are depicted in \textbf{Figure
  \ref{cl:fig4}} (results given for 10 clusters for both of user and
product clusters).

As you can see in the \textbf{Figure \ref{cl:fig4}}, \textbf{MCMMSBM}
optimizes entropy more efficiently than \textbf{MMSBM}. Moreover, we
can also set a trade-off: if we choose small value for $s$, for
instance $s$(=10), then \textbf{MCMMSBM} converges in the fastest way
but overall it converges too fast. For $s=30$, entropy is even smaller
than \textbf{MMSBM} at time 300sec. Thus, one can \emph{flexibly}
choose $s$ by considering hardware environment or time available. On
the other hand, if data size becomes huge, the time cost of one
iteration is very expensive in \textbf{MMSBM}. Moreover, if we choose
bigger $k, l$, then \textbf{MCMMSBM} becomes more efficient than 
\textbf{MMSBM}.

Now, let's evaluate the prediction. The measurment used is the RMSE
(Root Mean Square Error). For given pair of user and product in test
set, we can estimate the rating as \textbf{Eq (\ref{eq:estimate})}.
\begin{equation}
\label{eq:estimate}
\tilde{r}(u,v)=\sum_{r}\sum_{i,j}h_{u,i}\cdot h_{v,j}\cdot \theta_{i,j,r}\cdot r
\end{equation} 
Then, the RMSE is defined as \textbf{Eq (\ref{eq:rmse})}, where $Y$ 
is the test set.
\begin{equation}
\label{eq:rmse}
RMSE=\sqrt{\frac{1}{|Y|}\sum_{(u,v)\in Y}|r(u,v)-\tilde{r}(u,v)|^{2}}
\end{equation}

The result of RMSE is depicted as \textbf{Table \ref{cl:table2}}, we
can see that small entropy implies nice prediction.

\begin{table}
\centering
\begin{tabular}{| c | c | c |}
\hline 
 & Entropy & RMSE\\
\hline
\textbf{MMSBM} & 93876.4 & 0.9536\\
\hline
$s=10$ & 95952.2 & 0.9584 \\
\hline
$s=20$ & 93748.6 & 0.9515 \\
\hline
$s=30$ & 92920.0 & 0.9510 \\
\hline
\end{tabular}
\caption{\label{cl:table2}Running time: 300sec}
\end{table}
\subsection{Comparison between soft, and hard clustering}
In order to compare soft, and hard clustering, we also applied hard
clustering to recommender system. Here each user, and product belongs
to one of cluster \emph{deteministically}. The degree of freedom is
smaller than for soft clustering, thus the optimized entropy of training
set will be bigger. But, optimizing entropy is NOT our goal, don't
forget that our goal is \emph{prediction}. We can still expect better
prediction with hard clustering, despite of bigger entropy.

\begin{figure}[!ht]
\centering
\subfigure[]{\includegraphics[scale=0.4]{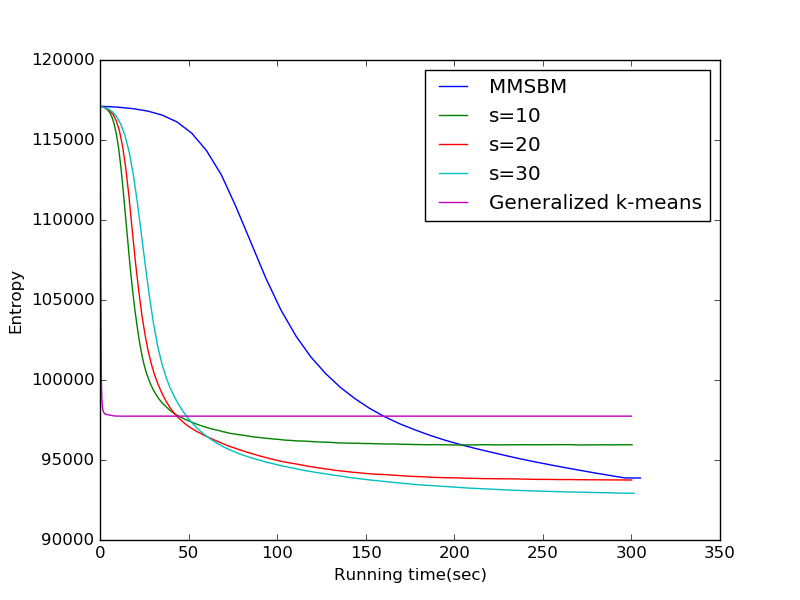}}
\subfigure[]{\includegraphics[scale=0.4]{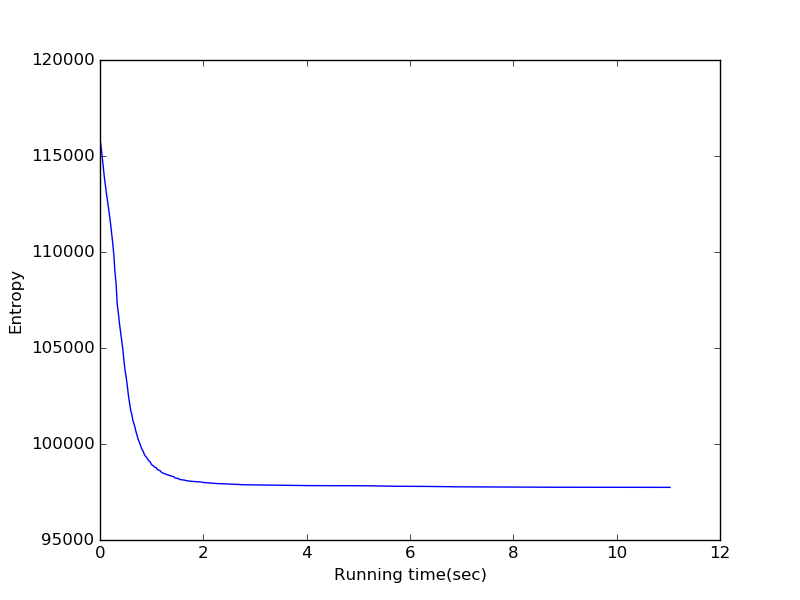}}
\caption{\label{cl:fig5}}
\end{figure}

First, let's compare the optimization of entropy. Results are
depicted in \textbf{Figure \ref{cl:fig5}}. The number of clusters is
set to 15 both for user and product clusters. We can see that hard
clustering (Generalized $k$-means) is much faster than soft
clustering, and that optimized entropy is bigger, as it was expected.

Next, let's evaluate prediction using RMSE. Results are depicted in
\textbf{Table \ref{cl:table3}}. The performance of prediction is also
worse than soft clustering. In summary: hard clustering
is cheaper, with a wrost quality than soft clustering.
\begin{table}[!ht]
\centering
\begin{tabular}{| c | c | c | c |}
\hline 
 & Entropy & RMSE & Running time\\
\hline
Generalized $k$-means  & 97741.0 & 0.9713 & 2 s\\
\hline
\textbf{MCMMSBM} ($s=30$) & 92920.0 & 0.9510 & 300 s \\
\hline
\end{tabular}
\caption{\label{cl:table3}}
\end{table}

\subsection{Result of social network anonymization}

Applying soft clustering to social network anonymization may be
untractable, because even for regenerating random network, it has a
complexity in $O(n^{2})$, while in hard clustering the complexitiy is
in $O(m)$, where $n$ is the number of vertices, and $m$ is the number
of edges (\emph{cf.} the complexity of generating ER random graph is
$O(n\cdot p)$). Thus here, we only apply hard graph clustering to
social network anonymization with Generalized $k$-means algorithm.

We used two networks from KONECT \cite{konect1} as benchmark of this
subsection. First network is Caida network. This is the undirected
network of autonomous systems of the Internet connected with each
other from the CAIDA project, collected in 2007. Nodes are autonomous
systems (AS), and edges denote communication ($|V(G)|=26,475,
|E(G)|=53,381$).

The second network is the arXiv astro-ph network. This is the collaboration
graph of authors of scientific papers from the arXiv's Astrophysics
(astro-ph) section. An edge between two authors represents a common
publication ($|V(G)|=18,771,|E(G)|=198,050$).

\begin{figure}
\subfigure[$\alpha=1.0$]{\includegraphics[scale=0.2]{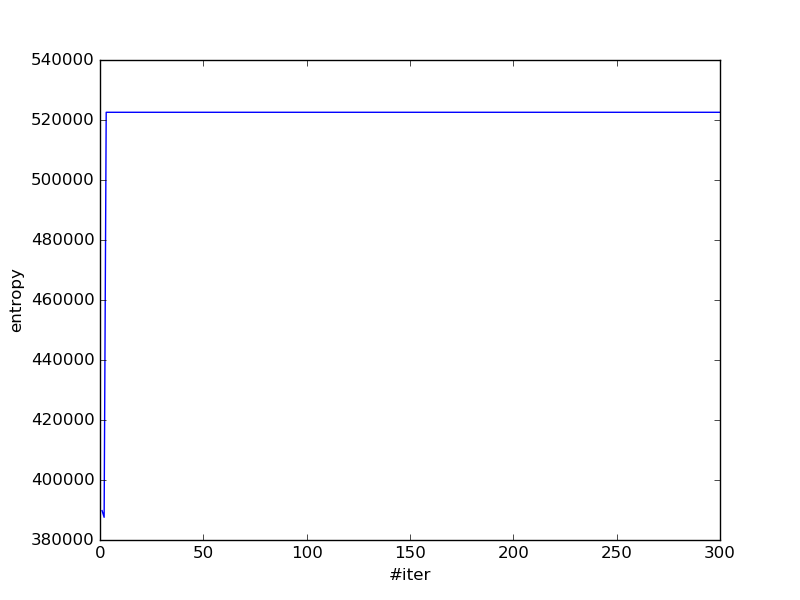}}
\subfigure[$\alpha=0.1$]{\includegraphics[scale=0.2]{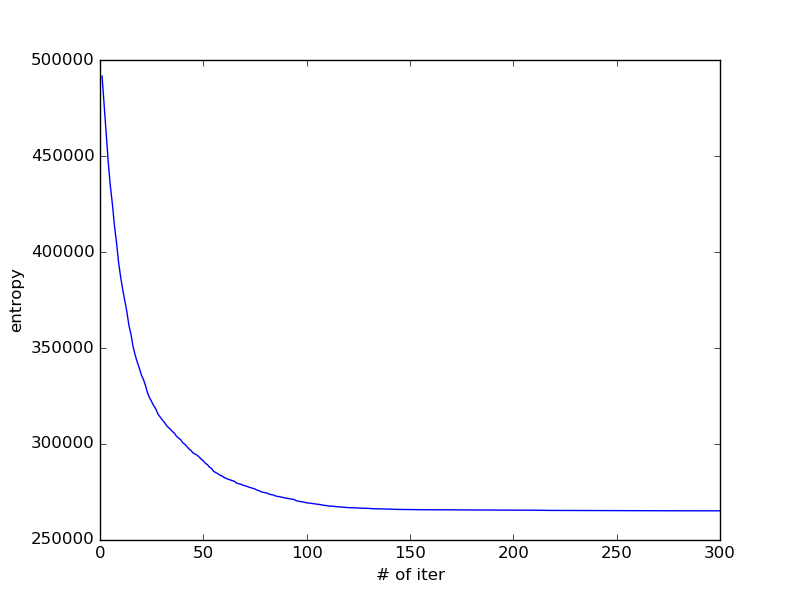}}
\caption{\label{fig:compare:ascaida}Caida network, 100-clustering, $\#$iter=300, Running time=10.6 s}
\end{figure}

\begin{figure}
\subfigure[$\alpha=1.0$]{\includegraphics[scale=0.2]{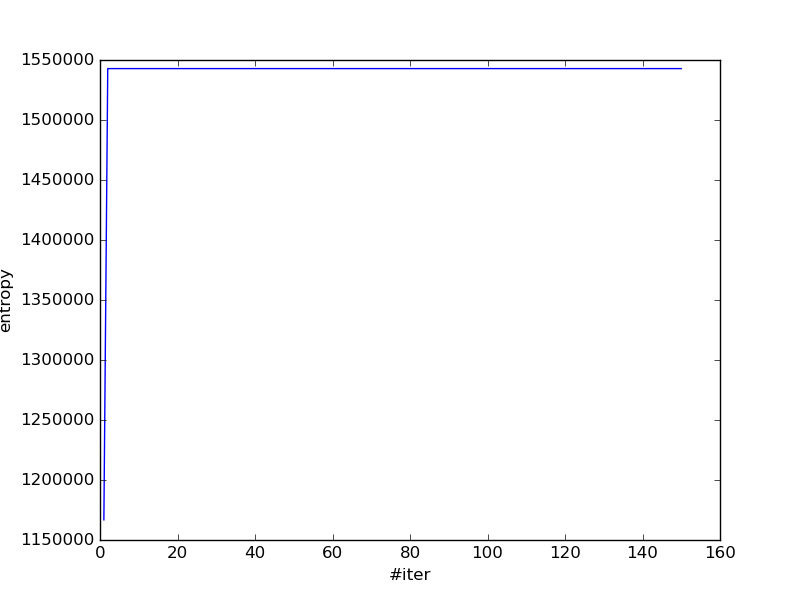}}
\subfigure[$\alpha=0.1$]{\includegraphics[scale=0.2]{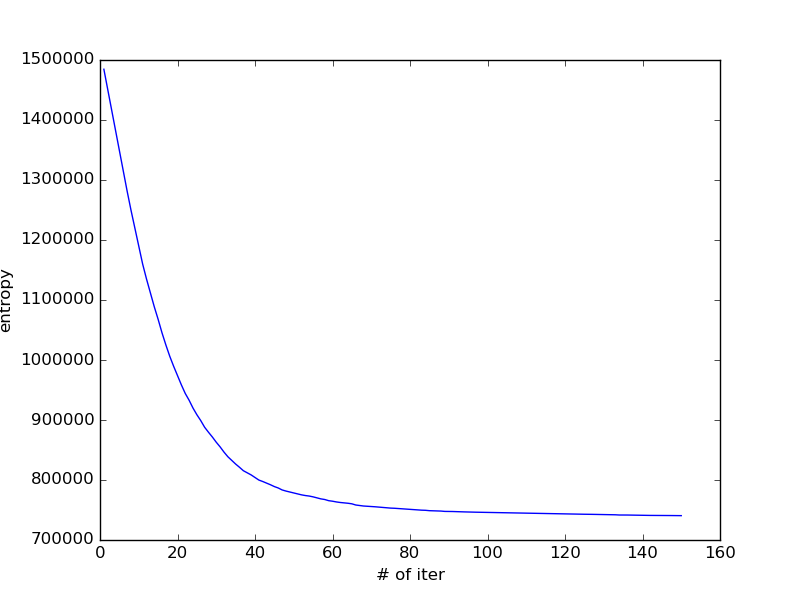}}
\caption{\label{fig:compare:astroph}arXiv astro-ph network, 300-clustering, $\#$iter=150, Running time=26 s}
\end{figure}

First, we show that using big iteration scale $\alpha$ doesn't work in
order to optimize entropy. Let us consider \textbf{Figure
  \ref{fig:compare:ascaida}} and \textbf{Figure
  \ref{fig:compare:astroph}}. We can see that using $\alpha=1.0$
doesn't work, as we argued previously, the distortion during the
parallel loop becomes an issue if we take a large $\alpha$. On the
other hand, when a small $\alpha$ is prefered (0.1), the distortion is
negligible, so that our algorithm optimizes entropy very well and
fast.

 
Next, we  evaluate the similarity between the original network and
the anonymized network. There are various measurements for evaluating
network similarity, here, we will compare APL (Average Path Length),
GCC (Global Clustering Coefficient), Degree distribution.

\begin{figure}[!ht]
\subfigure[Caida]{\includegraphics[scale=0.2]{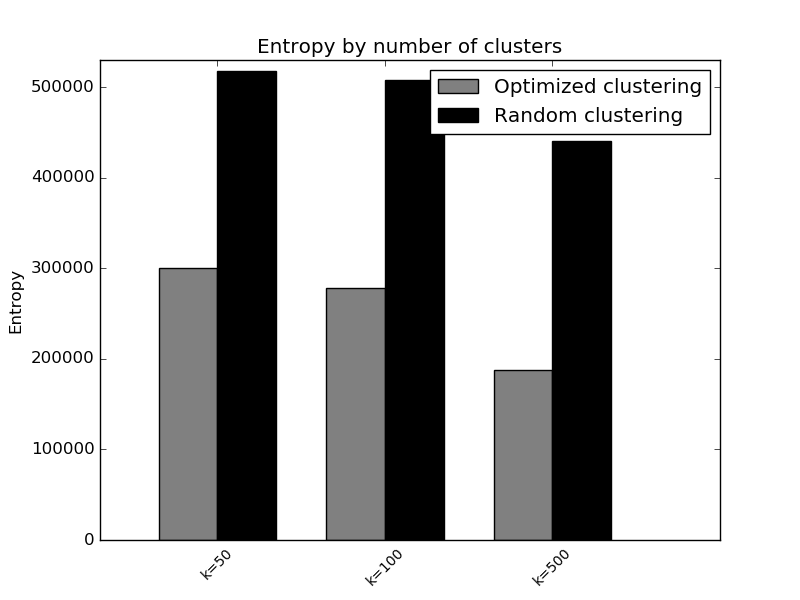}}
\subfigure[arXiv-AstroPh]{\includegraphics[scale=0.2]{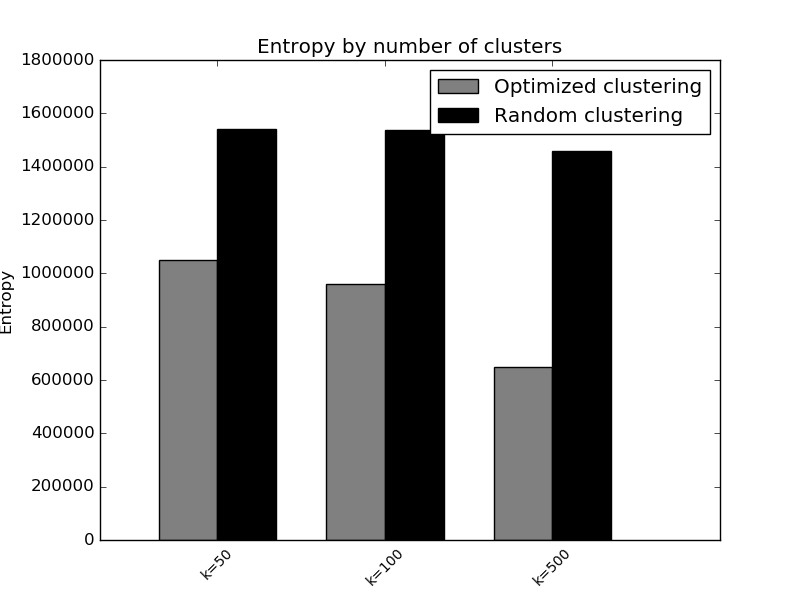}}
\caption{\label{cl:myfig1}Entropy}
\end{figure}

The similarity is compared by changing the number of clusters $k$. In
\textbf{Figure \ref{cl:myfig1}} are plotted the entropy of optimized
clustering and random clustering. We can see the number of clusters
doesn't affect much to entropy in random clustering, while in
optimized clustering, a larger number of clusters implies a better
entropy.

\begin{figure}[!ht]
\subfigure[Caida, optimized clustering]{\includegraphics[scale=0.2]{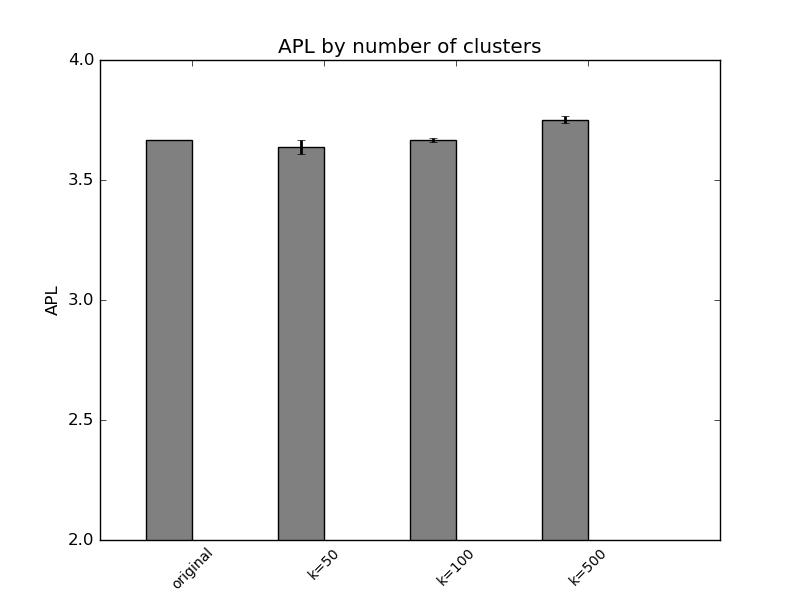}}
\subfigure[Caida, random clustering]{\includegraphics[scale=0.2]{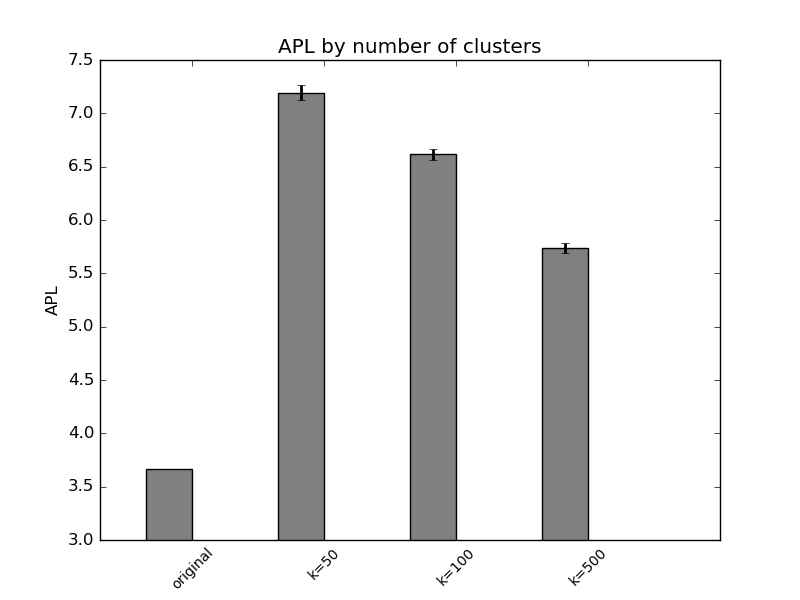}}
\subfigure[arXiv-AstroPh, optimized clustering]{\includegraphics[scale=0.2]{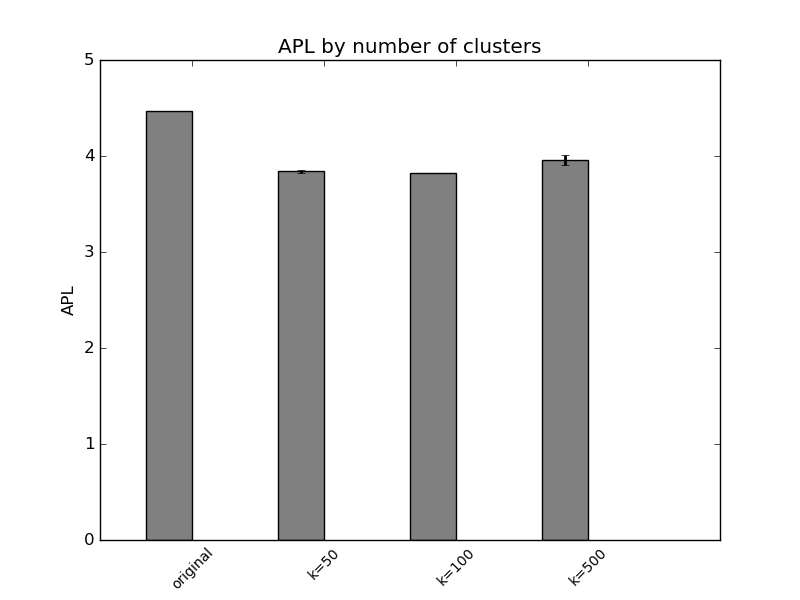}}
\subfigure[arXiv-AstroPh, random clustering]{\includegraphics[scale=0.2]{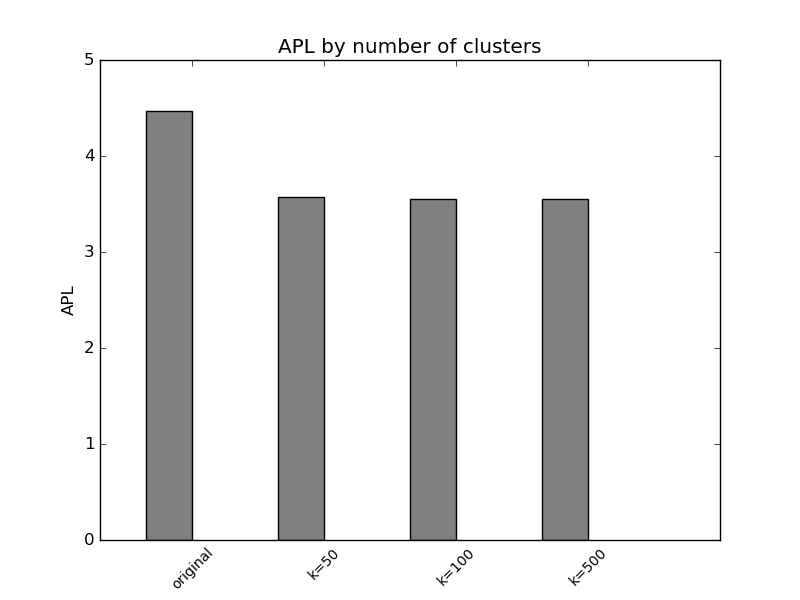}}
\caption{\label{cl:myfig2}APL}
\end{figure}

\begin{figure}[!ht]
\subfigure[Caida, optimized clustering]{\includegraphics[scale=0.2]{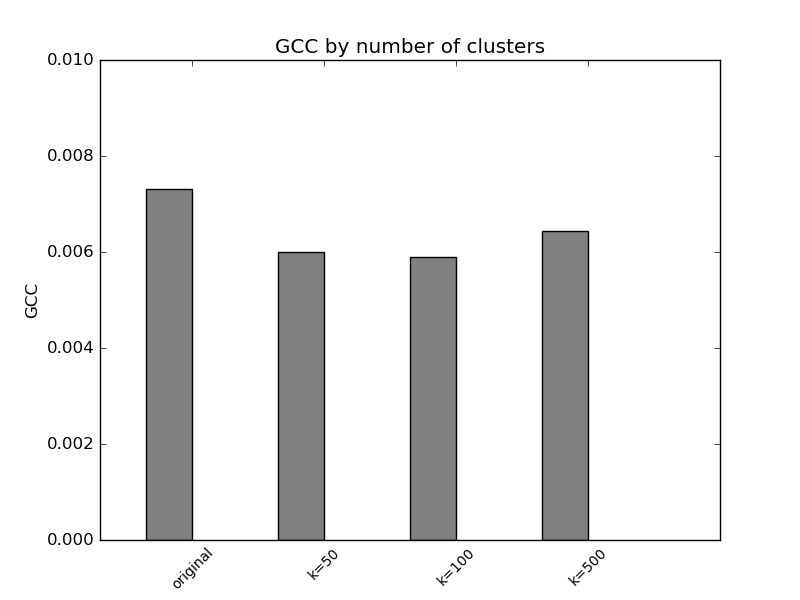}}
\subfigure[Caida, random clustering]{\includegraphics[scale=0.2]{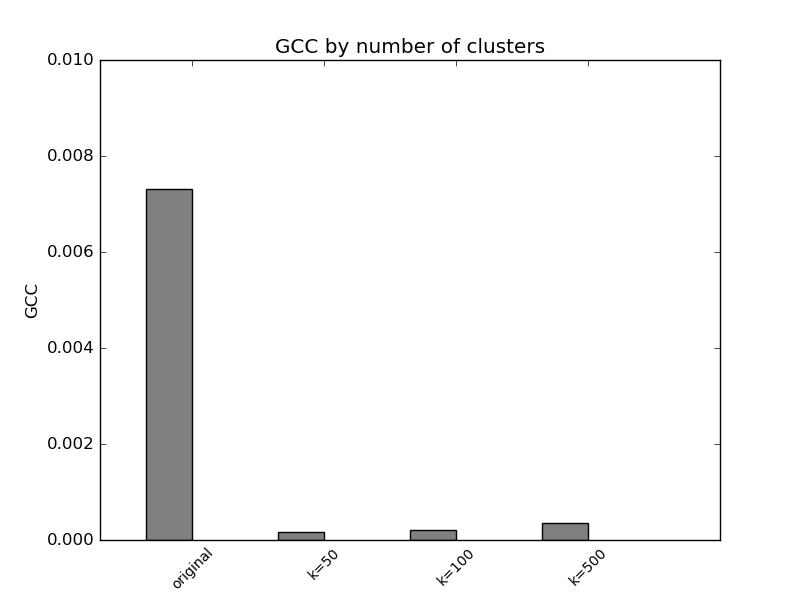}}
\subfigure[arXiv-AstroPh, optimized clustering]{\includegraphics[scale=0.2]{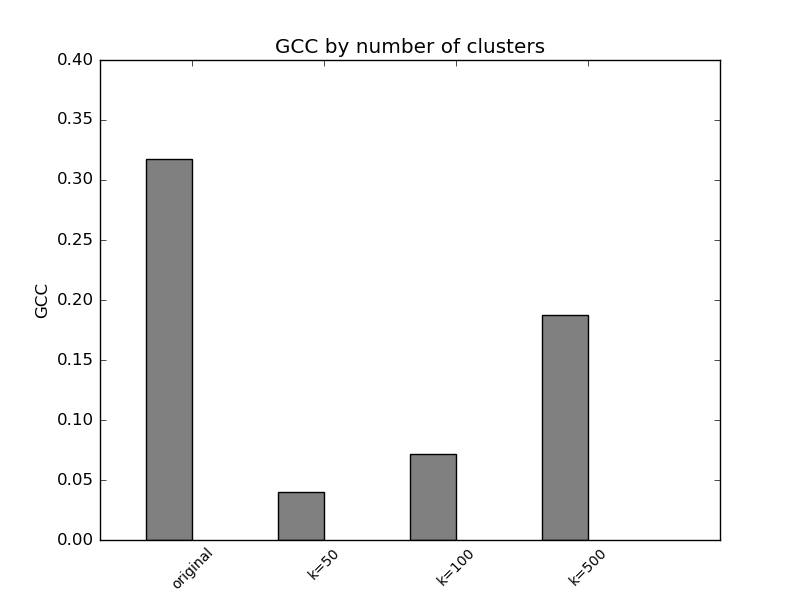}}
\subfigure[arXiv-AstroPh, random clustering]{\includegraphics[scale=0.2]{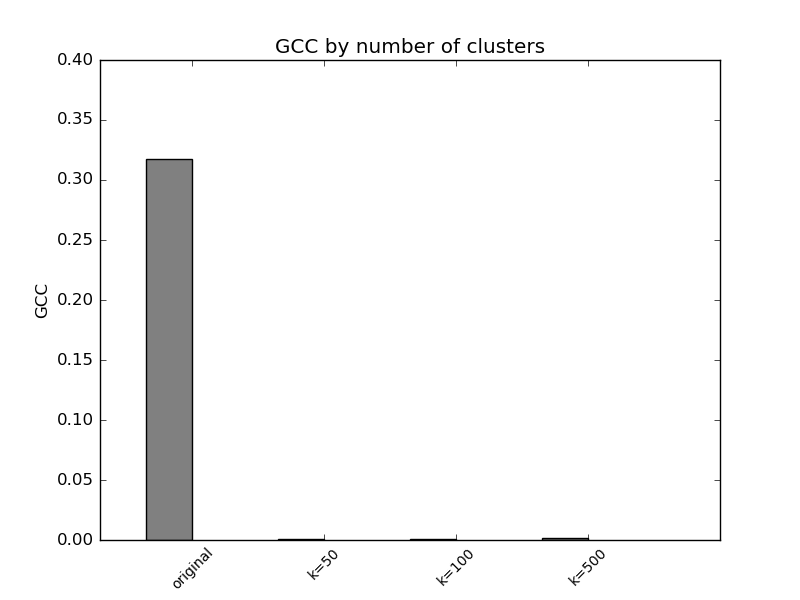}}
\caption{\label{cl:myfig3}GCC}
\end{figure}

Now let's see how this entropy affect the general properties of the
graphs. Since anonymized network are \emph{randomly} generated, we
have generated them 5 times, and then estimated APL, GCC for each
network. Thus the error bar is also represented in \textbf{Figure
  \ref{cl:myfig2}, \ref{cl:myfig3}}, but we can see that such error
bars are very tight.

\begin{figure}[!ht]
\subfigure[Caida, optimized clustering]{\includegraphics[scale=0.2]{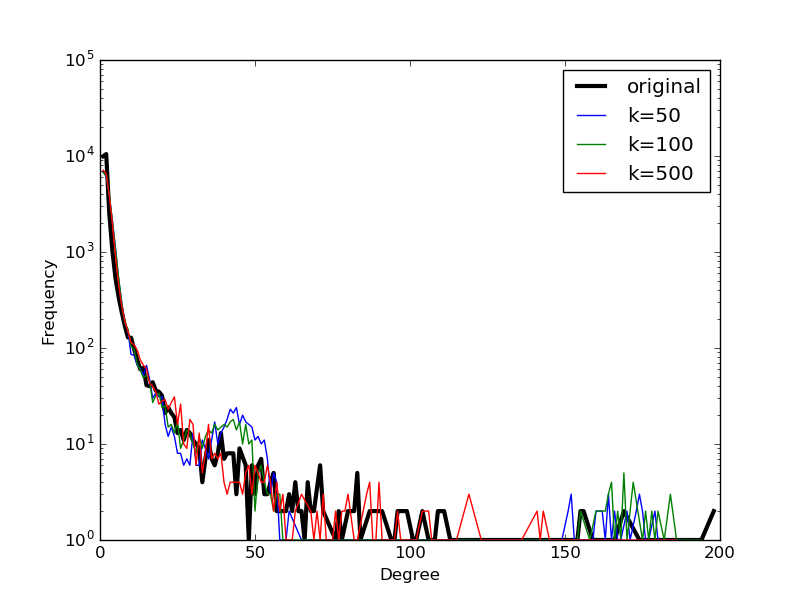}}
\subfigure[Caida, random clustering]{\includegraphics[scale=0.2]{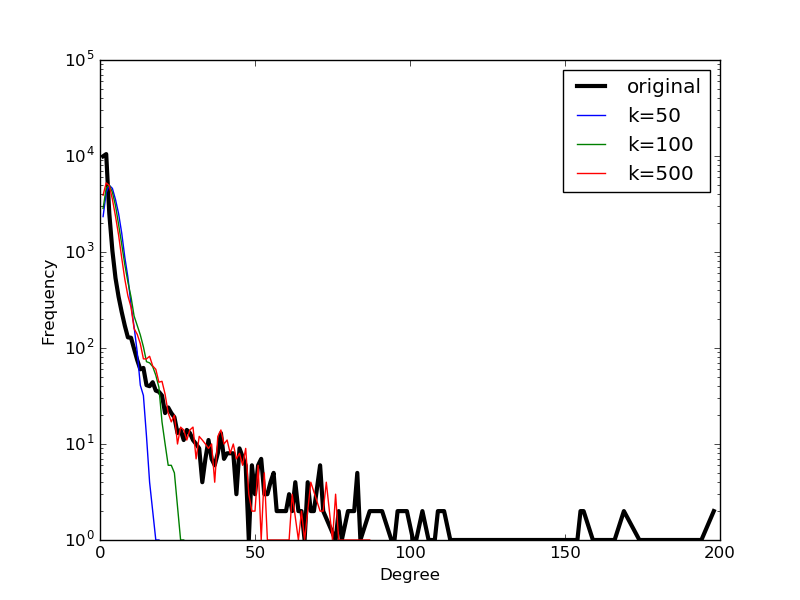}}
\subfigure[arXiv-AstroPh, optimized clustering]{\includegraphics[scale=0.2]{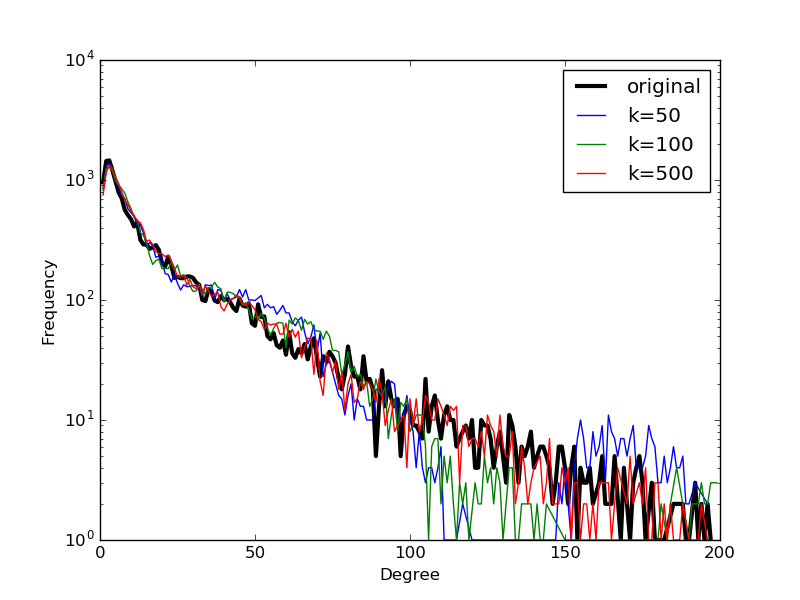}}
\subfigure[arXiv-AstroPh, random clustering]{\includegraphics[scale=0.2]{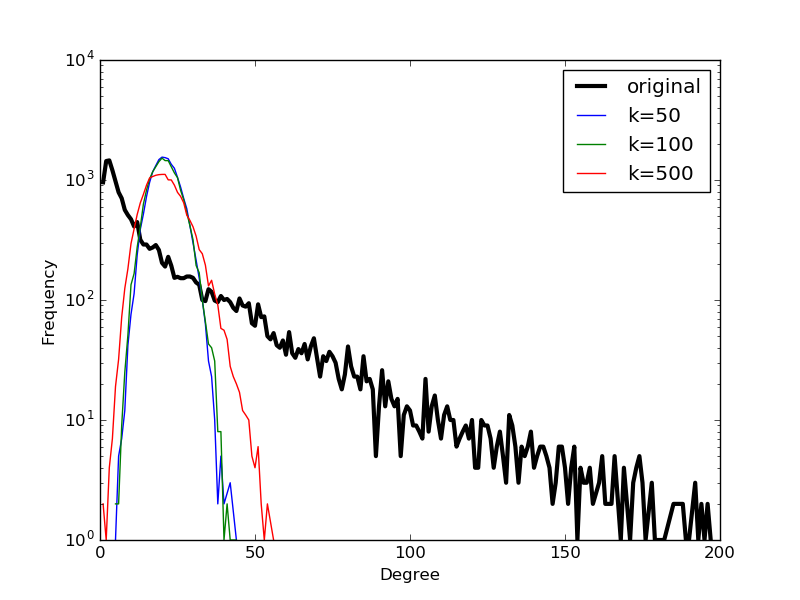}}
\caption{\label{cl:myfig4}Degree distribution}
\end{figure}

\section{Related Works}
\label{sec:related_works}


  \textbf{MCMMSBM (Algorithm \ref{cl:alg4})} is closely related to the 
\textbf{MMSBM} (originally defined in
\cite{ABFX08}) and more precisely to the algorithm of \cite{GLGMS16}:
see \textbf{Algorithm \ref{cl:alg3}}. 

\begin{algorithm}[!ht]
\SetKwInOut{Input}{Input}
\SetKwInOut{Output}{Output}
\Input{observed edge data $X$, $V(G)$: set of vertices, $k$: number of clusters }
\Output{optimal $\theta,h$}
Take random $\hat{\theta},\hat{h}$\\
\Repeat{$S(\theta,h)$ converges}{
\ForPar{$(u,v)\in X$, and $i,j\in [k]$}{
$x_{u,v}(i,j)\gets \frac{h_{u,i}\cdot h_{v,j} \cdot \theta_{i,j,r(u,v)}}{\sum_{i',j'\in K}h_{u,i'}\cdot h_{v,j'}\cdot \theta_{i',j',r(u,v)}}$
}
\ForPar{$u\in V(G),i\in [k]$}{
$h_{u,i}\gets \frac{1}{|\partial u|}\cdot \sum_{v\in \partial u}\sum_{j\in K}x_{u,v}(i,j)$
}
\ForPar{$i,j\in [k],r\in R$}{
$\theta_{i,j,r}\gets \frac{\sum_{(u,v)\in X,r(u,v)=r}x_{u,v}(i,j)}{\sum_{(u,v)\in X}x_{u,v}(i,j)}$
}
}
\caption{\label{cl:alg3}Mixed Memebership SBM(MMSBM)}
\end{algorithm}

Note that lines 6$\sim$8 correspond to
the M-step of EM algorithm, instead of the E-step. Actually, the lines
3$\sim$5 correspond to the E-step, and lines 6$\sim$11 correspond to
the M-step. The time complexity of algorithm \ref{cl:alg3} is
$O(|X|\cdot k^{2})$. If $k$ is considered as constant, the complexity
is thus $O(|X|)$. However, if the size of data become huge, then the
number of clusters $k$ may also increase. Moreover, $k^{2}$ is
\emph{never} negligible, even if $k$ is small.

On the other hand, the time complexity of \textbf{MCMMSBM} is
$O(|X|\cdot (k +s\cdot log(k) ))$, the sample size $s$ being chosen by
trade-off between cost and accuracy.

The update of $h_{u,i}$ in \textbf{Algorithm \ref{cl:alg3}} (line 7) is
computed as follows:

\begin{equation}
\begin{split}
h_{u,i}=
  &\frac{1}{|\partial u|}\cdot \sum_{v\in \partial u}\sum_{j\in K}x_{u,v}(i,j)\\
 =&\frac{1}{|\partial u|}\cdot \sum_{v\in \partial u}
   \frac{\underset{j'\in K}{\sum} h_{u,i}\cdot  h_{v,j'}\cdot 
   \theta_{i,j',r(u,v)}}{\underset{i',j'\in K}{\sum}h_{u,i'}
   \cdot h_{v,j'}\cdot \theta_{i',j',r(u,v)}}
\end{split}
\end{equation}

The computation of the denominator requires a computation in $O(k^{2})$.

If $h_{u,i'}\cdot h_{v,j'}$ is  considered as a pmf of some random
variable we have to define the following random variables:

\begin{equation}
\begin{split}
X_{u,v}   
   &:=\theta_{i',j',r(u,v)} \text{ w/p $h_{u,i'}\cdot h_{v,j'}$}\\
X_{u,v}(i)
  &:=\begin{cases}
    \theta_{i,j',r(u,v)} &\text{w/p $h_{u,i}\cdot h_{v,j'}$}\\
    0 &\text{w/p $1-h_{u,i}$}
  \end{cases}\\
X_{u,v}(i,j)
  &:=\begin{cases}
    \theta_{i,j,r(u,v)} &\text{w/p $h_{u,i}\cdot h_{v,j}$}\\
    0 &\text{w/p $1-h_{u,i}\cdot h_{v,j}$}
  \end{cases}
\end{split}
\end{equation} 
where w/p means ``with probability", and $i',j' \in [k]$, then 

\begin{equation}
\begin{split}
\underset{i',j'\in [k]}
         {\sum}h_{u,i'}\cdot h_{v,j'}\cdot \theta_{i',j',r(u,v)}
  &=\mathbb{E}[X_{u,v}]\\
\underset{j'\in [k]}{\sum}h_{u,i}\cdot h_{v,j'}\cdot
          \theta_{i,j',r(u,v)}
  &=\mathbb{E}[X_{u,v}(i)]
\end{split}
\end{equation}

$X_{u,v},X_{u,v}(i),X_{u,v}(i,j)$ can be generated as follows:

\begin{enumerate}
\item{Pick $i',j'$ respectively from $discrete(h_{u}),discrete(h_{v})$}
\item{Assign the value of random variables $X_{u,v},X_{u,v}(i),
     X_{u,v}(i,j)$ as: \begin{equation}
\begin{split}
  X_{u,v} & 
    :=\theta_{i',j',r(u,v)}\\ 
  X_{u,v}(i) &
    :=\begin{cases}
        \theta_{i',j',r(u,v)} &\text{if $i'=i$}\\
        0 &\text{else}
    \end{cases}\\
  X_{u,v}(i,j)&
    :=\begin{cases}
      \theta_{i',j',r(u,v)} &\text{if $i'=i$ and $j'=j$}\\
      0 &\text{else}
    \end{cases}
\end{split}
\end{equation}}
\end{enumerate}

\noindent
Those random variables can be sampled in $O(k+s\cdot log(k))$ for
sample size $s$. Indeed, it is a joint distribution of 
\emph{independent} events. For $s'\in [s]$, we can sample 

\begin{equation}
\begin{split}
  X_{u,v,s'}     &\stackrel{i.i.d}{\sim}X_{u,v}\\
  X_{u,v,s'}(i)  &\stackrel{i.i.d}{\sim}X_{u,v}(i)\\
  X_{u,v,s'}(i,j)&\stackrel{i.i.d}{\sim}X_{u,v}(i,j)
\end{split}
\end{equation} 

\noindent
Thus through Monte-Carlo simulation, $g,\theta$ can be approximately
updated as follows: 

\begin{equation}
\begin{split}
  h_{u,i} & =\frac{1}{|\partial u|}
             \sum_{v\in \partial u}
             \frac{\mathbb{E}[X_{u,v}(i)]}{\mathbb{E}[X_{u,v}]}\\
         &\approx \frac{1}{|\partial u|}
             \sum_{v\in \partial u}
             \frac{\frac{1}{s}\underset{s'}{\sum}X_{u,v,s'}(i)}
                  {\underset{s'}{\sum}X_{u,v,s'}}\\
  \theta_{i,j,r}&=\frac{\underset{(u,v)\in
                       X,r(u,v)=r}{\sum}x_{u,v}(i,j)}
                     {\underset{(u,v)\in X}{\sum}x_{u,v}(i,j)}\\
          &\approx \frac{\underset{(u,v)\in
              X,r(u,v)=r}{\sum}\tilde{x}_{u,v}(i,j)}
              {\underset{(u,v)\in X}{\sum}\tilde{x}_{u,v}(i,j)}
\end{split}
\end{equation} 

\noindent
where 

\begin{equation}
  \tilde{x}_{u,v}(i,j):=
  \frac{\frac{1}{s}\underset{s'}{\sum}X_{u,v,s'}(i,j)}
       {\frac{1}{s}\underset{s'}{\sum}X_{u,v,s'}}
\end{equation}

 It is computed in $O(|X|\cdot (k+s\cdot log(k)))$ time by 
\textbf{Algorithm \ref{cl:alg4}}.

\medskip
The application of the Generalized $k$-means algorithm to the problem of social network
anonymization amounts to an approach similar to the one of proposed in
\cite{hay10} by Hay \emph{et al}. We can note the
following differences between two approaches are:
\begin{enumerate}
    \item{In \cite{hay10} the edges between two clusters are just
    rearranged, thus the number of edges between two clusters is a constant.
    With our approach the number may change.}

    \item{With our approach, a totally new random graph is generated which is
    only \emph{similar} to the original one. The concept of
    $k$-anonymity cannot be strictly applied. Namely we don't have to fix
    the minimum size of clusters which differs with \cite{hay10}.} \smallskip
\end{enumerate}
Moreover, the experimental results shows a very large difference in
terms of efficiency between our approach and \cite{hay10}. Despite
different hardware configuration with \cite{hay10}, Generalized
$k$-means appears much faster. For example in \cite{hay10} the
following result is given: it takes 1 hour to cluster graphs of size
5000, on the other hand it takes 10 seconds using Generalized
$k$-means to cluster graphs of size 30000.

\section{Conclusion}
\label{sec:conclusion}
In this paper we have presented the \textbf{MCMMSBM} algorithm which
can be seen as an improved \textbf{MMSBM} algorithm \cite{GLGMS16} by
applying Monte-Carlo simulation to point of
\emph{efficiency}. Theoretically, \textbf{MCMMSBM} can't achieve
better optimizations than \textbf{MMSBM}, if one considers
\emph{infinite} computing resources. But in reality,  \textbf{MMSBM} is
\emph{strictly} limited. We have shown that \textbf{MCMMSBM} can achieve
better optimization, and better prediction in \textbf{Section
  \ref{cl:sec6.1}}.

We also proposed the Generalized $k$-means algorithm. It can be widely
applied for hard classification problems, especially for hard graph
clustering problems. We reclassified small proportion of data (or
nodes), instead of the whole data in one iterration.

We have applied SBM to social network anonymization. We saw that
entropy optimization works very well for property preservation. We
also compared the results by changing the number of clusters $k$, we
can consider that there can be trade-off for deciding $k$. If we take
small $k$, we saw that network properties are not preserved well. But
if we take large $k$, \emph{anonymity} can be vulnerable. We left as
future work deeper comparisons with other social network
anonymization techniques, notably on the quality of the published
network, for instance with \cite{TrutaCR12,CampanAT15}. 

In order to measure the anonymity achieved, we have relied on the
$k$-anonymity \cite{BackstromDK11} definition which is widely
applied. But such definition is not really suitted for this
anonymization framework. As future work we consider  
to define an appropriate anonymity measurement for this 
framework.


\balance
\bibliographystyle{abbrv}

\end{document}